\documentclass[preprint,aps,pra,showpacs]{revtex4}
\usepackage{tabularx}
\usepackage{dcolumn}
\usepackage{graphics}
\usepackage{bm}
\usepackage[dvips]{graphicx}
\usepackage[dvips]{rotating}
\usepackage[latin1]{inputenc}
\usepackage{dcolumn}
\usepackage{tabularx}
\usepackage{amsmath}
\usepackage{amssymb}
\begin{document}
\draft

\title{Spatial oscillations in the spontaneous emission rate of an atom inside a metallic wedge}
\author{H. J. Zhao$^{1}$ and  M. L. Du$^2$ }\email{
duml@itp.ac.cn} \affiliation{$^{1}$School of Physics and Information
Science, Shanxi Normal University, Linfen 041004, China}
 \affiliation{$^2$Institute
of Theoretical Physics, Chinese Academy of Sciences, P. O. Box 2735,
Beijing 100190, China}

\date{\today}

\begin{abstract}
A method of images is applied to study the spontaneous emission of an atom inside a metallic wedge with an opening angle of $\pi/N$, where $N$ is an arbitrary positive integer. We show the method of images gives a rate formula consistent with that from Quantum Electrodynamics. Using the method of images, we show the correspondence between the oscillations in the spontaneous emission rate and the closed-orbits of emitted photon going away and returning to the atom inside the wedge. The closed-orbits can be readily constructed using the method of images and they are also extracted from the spontaneous emission rate.
\par

\pacs{37.30.+i,32.70.Cs, 12.20.Ds}

\par
\keywords {spontaneous emission rate, metallic wedge, closed-orbits}

\end{abstract}

 \maketitle


\section{Introduction}
The effect of boundaries and environments on spontaneous emission
of atoms is an interesting subject in the physics of radiation with a long history \cite{Purcell}. Our understanding of the subject has been greatly improved by the theoretical and experimental developments over the years \cite{Milonnipaper,Buzek,Urbach,Kleppner,Goy}. For example, Hulet \emph{et al} in an experiment demonstrated clearly that the lifetime of atomic state can be greatly modified by the two metal plates \cite{Huletetal}. From a practical point of view, once the boundaries are given and specified for a system, we would like to know how the spontaneous emission rate is modified. In particular, how does the spontaneous emission rate depends upon the position and polarization of the emitting atom in the system. Quantum Electrodynamics (QED) relates the spontaneous emission rate of an atom to the quantized electromagnetic
field \cite{Milonnibook,Hind}.
In principle, the calculations of the spontaneous emission rate of an atom in a cavity become therefore the search for the quantized electromagnetic
filed of the specified system, the evaluation of the matrix elements and the summation of the matrix elements squared. In practice, however, the above process can be complicated and difficult to apply even for a simple system such as a metallic wedge \cite{Rosa,Skipsey}. Therefore there is a need for alternative methods.

In this article we will apply a theoretical method of images to study the spontaneous emission rate in a metallic wedge. In an interesting application, the method of images has recently been used to study the quantum interference at corners\cite{Guze}.
We will show the method works for a wedge with an angle of $\pi/N$, where $N$ is an arbitrary positive integer. As examples, we have worked out the details for wedges of $\pi/3$ and $\pi/4$ corresponding to $N=3$ and $N=4$.
We find the method of images gives a formula for the spontaneous emission rate which is consistent with the one obtained by Rosa \emph{et al} involving long derivations using QED. The method of images is not only easy to apply, it also provide a way to understand the rate formula by associating each term with an image and a closed-orbit. For other systems, we have shown the large scale oscillations in the spontaneous emission rate \cite{DuSE1,DuSE2} are similar to the oscillations for photo-ionization and photo-detachment in the presence of external fields \cite{DuCOT1,DuCOT2,DuCOT3,DuCOT4,kleppnerdelos}. Using scaling variables, we have  demonstrated the direct correspondence between the oscillations in the spontaneous emission rate and the closed-orbits of photon \cite{DuSE1,DuSE2}. Here we will use the same method to analyze the oscillations in the spontaneous emission rate in a wedge with an angle of $\pi/3$ and $\pi/4$ respectively. We will show there are four distinct oscillations for a $\pi/3$ wedge, and there are six distinct oscillations for a $\pi/4$ wedge.

The paper is organized as follows. In the next section, we discuss the construction of the images and the calculation of the emission rate. The method of images is general in the sense that it can be applied to a wedge with an opening angle of $\pi/N$, where $N$ is an arbitrary positive integer. In section III, we take a $\pi/3$ wedge and a $\pi/4$ wedge as examples and discuss the oscillation patterns of the spontaneous emission rate as the position of the emitting atom inside the wedge is changed. We then analyze and extract the oscillations using transformations. We will demonstrate the direct correspondence between the oscillations and photon closed-orbits inside the wedges. Conclusions and perspectives will be given in section IV.

\section{the method of images and the emission rate}

The cross sectional view of a metallic wedge is shown schematically in Fig.1. The wedge is assumed to be made of perfect metal and an atom is placed inside the wedge at the position of the black dot. Let $\theta$ be the opening angle of the wedge. Let the z-axis be the wedge axis. A cylindrical coordinate system ($\rho,\phi,z$) is defined such that the distance between the atom and the z-axis axis is $\rho$ and the declination angle relative to the downside plane is $\phi$.
A Cartesian coordinate system ($x,y,z$) is also useful. We define the x-axis in the bisector plane of the opening and the y-axis perpendicular to it. For the sake of simplicity, the dipole moment of the atom in the following is assumed to be parallel to the wedge axis.

The method of images has been applied to discuss the spontaneous emission rate of an atom near a mirror and between two parallel plates \cite{Hind}. The analogy of an radiating antenna to an atomic emission process is explored
in the method of images.  For an emitting atom near a plane mirror, it has been demonstrated that the spontaneous emission rate can be inferred from
the power radiation rate of an antenna \cite{Hind}.
Consider the radiation damping of a dipole
antenna with the dipole moment $\vec{d}e^{-i\omega_0 t}$ , where
$\omega_0$ is the frequency. The radiation damping
rate is defined as $W_d=P/U$, where $P$ is the average radiation power of the antenna and $U$ is the mechanical energy of the antenna. Let $\vec{r}$ denote a vector of a point relative to the dipole position,the radiation damping rate can be written as \cite{Hind}
\begin{equation}\label{1}
  W_d=\frac{\omega_0}{2U}{\rm{Im}}[\vec{d}^*\cdot\vec{E}(\vec{r}=0)]
  =\frac{\omega_0}{2U}\{{\rm{Im}}[\vec{d}^*\cdot\vec{E}_0(\vec{r}=0)]
  +{\rm{Im}}[\vec{d}^*\cdot\vec{E}_{ref}(\vec{r}=0)]\},
\end{equation}
where the electric field  $\vec{E}$ at the
position of dipole antenna has been decomposed into a direct part $\vec{E}_0$  and a reflected part $\vec{E}_{ref}$. The direct
electric field is
\begin{equation}\label{2}
   \vec{E}_0=\frac{dk^3}{4\pi \varepsilon_0}\{(\hat{r}\times\hat{d})\times\hat{r}(\frac{1}{kr})
   +[\hat{d}-3\hat{r}(\hat{r}\cdot\hat{d})]
   \times(\frac{i}{(kr)^2}-\frac{1}{(kr)^3})\}e^{i(kr-\omega_0t)},
\end{equation}
where $\hat{d}$ is a unit vector in the dipole direction and $\varepsilon_0$ is the vacuum permittivity.
If the dipole antenna is at distance
$l/2$ from a plane metal boundary and is parallel to the metal surface, the effect of the conducting boundary can be replaced by an image dipole and the reflected electric field $\vec{E}_{ref}$ of the dipole can be calculated directly from the imaging dipole. The imaging dipole direction is opposite to the original dipole direction and is located at a distance $l/2$ behind the metal surface.
The reflected field at the position of the antenna is
\begin{equation}\label{3}
    \vec{E}_{ret}=-\frac{dk^3}{4\pi \varepsilon_0}\hat{d}
    (\frac{1}{kl}+\frac{i}{(kl)^2}
    -\frac{1}{(kl)^3})e^{i(kl-\omega_0t)},
\end{equation}
where $l$ is
the distance between the image and the original antenna.

When a dipole antenna is placed in a metallic wedge as shown in Fig.1,
the reflected field can be calculated by using a set of images and the reflected filed is a sum of contributions
from each image of the set. We find when the opening angle of the wedge is $\theta=\pi/N$, where $N$ is an arbitrary positive integer, the number of the images is finite. The images can be constructed using a technique derived from a wave-packet evolution study inside a wedge by Robinett \cite{Robinett}. If the emitting atom's cylindrical coordination is at $(\rho, \phi,z)$, the first image with a dipole $-\vec{d}$ will appear at $(\rho, -\phi,z)$. The original source and the first image can be rotated
by $ 2\theta $ clockwise to give the second image located at $(\rho, \phi-2\theta,z)$ with a dipole $\vec{d}$ and the third image at $(\rho, -\phi-2\theta,z)$ with a dipole $-\vec{d}$. The original source and the first image are rotated by $4\theta$ clockwise to give image 4 located at $(\rho, \phi-4\theta,z)$ with a dipole $\vec{d}$ and image 5 at $(\rho, -\phi-4\theta,z)$ with a dipole $-\vec{d}$. This process can be continued until the last two images are obtained by rotating the original source and the first image by $ 2(N-1)\theta $ clockwise. For $\theta=\pi/N$, where $N$ is an arbitrary positive integer, this rotation process closes on itself. In the end, we have one original source and $(2N-1)$ images. $(N-1)$ of those images with the same dipole $\vec{d}$ as the original's are located at $(\rho,\phi-2n\pi/N,z)$, where $n=1,2,\ldots,(N-1)$. The other $N$ images are located at $(\rho,-\phi+2n\pi/N,z)$, where $n=0,1,\ldots, (N-1)$, and they have the opposite dipole $-\vec{d}$ as the original's. Fig.2 shows the five images and the original source for a $\theta=\pi/3$ wedge.

The reflected electric field at the position of the source antenna in  the wedge with an opening angle $\theta=\pi/N$ can be obtained from the  electric fields of all the images using Eq.(3). The distance between the source at $(\rho,\phi,z)$ and the image at $(\rho,-\phi,z)$ is $l_-=2\rho\sin (\phi)$; the distance between the source at $(\rho, \phi,z)$ and the images at $(\rho,\phi-2n\pi/N,z)$, where $n=1,2,\ldots,(N-1)$, are $l_{n+}=2\rho\sin (n\pi/N)$; the distance between the source at $(\rho,\phi,z)$ and $(\rho, -\phi+2n\pi/N,z)$, where $n=1,2,\ldots,(N-1)$, are $l_{n-}=2\rho\sin (n\pi/N-\phi)$.

The reflected field at the position of the original dipole antenna can
be written as
\begin{eqnarray}\label{4}
  \nonumber  \vec{E}_{ref}&=&-\frac{dk^3}{4\pi \epsilon_0}\hat{d}
   [\frac{1}{kl_-}+\frac{i}{(kl_-)^2}-\frac{1}{(kl_-)^3}]e^{i(kl_--\omega_0t)}\\
   \nonumber  & -&\frac{dk^3}{4\pi \epsilon_0}\hat{d}\sum_{n=1}^{N-1}
   \{[\frac{1}{kl_{n-}}+\frac{i}{(kl_{n-})^2}-\frac{1}{(kl_{n-})^3}]e^{i(kl_{n-}-\omega_0t)}\\
   &-&[\frac{1}{kl_{n+}}+\frac{i}{(kl_{n+})^2}-\frac{1}{(kl_{n+})^3}]e^{i(kl_{n+}-\omega_0t)}\}.
\end{eqnarray}
Inserting Eq. (\ref{4}) and Eq. (\ref{2}) into Eq. (\ref{1}) and after some manipulations, we get the damping rate inside the wedge
\begin{equation}\label{5}
  W_d=W_0\{1-\frac{3}{2}G_{\|}[2k\rho\sin \phi]
  -\frac{3}{2}\sum_{n=1}^{N-1}[G_{\|}[2k\rho\sin(\frac{n\pi}{N}-\phi)]
  -G_{\|}[2k\rho\sin(\frac{n\pi}{N})]]\},
\end{equation}
where we have defined
\begin{equation}\label{6}
    G_{\|}(S)=\frac{\sin S}{S}
    +\frac{\cos S}{S^2}-\frac{\sin S}{S^3}
\end{equation}
and
$W_0=\frac{2}{3}\frac{|d|^2k^3}{4\pi\epsilon_0}\frac{\omega_0}{2U}$ is
the damping rate in free space \cite{Hind}.

The spontaneous emission rate in a wedge $\gamma$ can be inferred from Eq.(5) as
\begin{equation}
  \gamma=\gamma_0\{1-\frac{3}{2}G_{\|}[2k\rho\sin \phi]
  -\frac{3}{2}\sum_{n=1}^{N-1}[G_{\|}[2k\rho\sin(\frac{n\pi}{N}-\phi)]
  -G_{\|}[2k\rho\sin(\frac{n\pi}{N})]]\},
\end{equation}
where $\gamma_0$ is the spontaneous emission rate without the wedge.
It should be noted that the method of images has its limitation and the correct value of $\gamma_0$ from QED will have to be used in Eq.(7)\cite{Hind,Meschede}.

Rosa \emph{et al} studied the emitted power of an atom
in a perfect wedge with an arbitrary opening angle using QED. The emitted power is proportional to the spontaneous emission rate. The formula derived by Rosa \emph{et al} using QED involves special functions and appears quite complicated. But when the opening angle of the wedge is restricted to $\theta=\pi/N$, where $N$ is an arbitrary positive integer, the QED formula can be simplified considerably (see Eq. (46) in Ref. \cite{Rosa}).
We find Eq.(7) using the method of images is consistent with the result of
Rosa \emph{et al} using QED for a $\theta=\pi/N$ wedge. It is not easy to see why the complicated QED formula involving special functions obtained by Rosa \emph{et al} can be reduced to a simple form similar to Eq.(7) when the opening angle of the wedge is $\theta=\pi/N$, where $N$ is an arbitrary positive integer. The method of images seems to suggest that the origin of the simplification is in the symmetries of the finite number of images. When the opening angle is $\pi/N$, where $N$ is an arbitrary positive integer, the $2N-1$ images and the original source form a set, this set is invariant when it is rotated by a multiple of $2\pi/N$ either clockwise or anti-clockwise. Furthermore, the set can be changed to a another set by an inversion through either the upper plane or the lower plane of the wedge. The effect of the inversion is equivalent to reversing the dipole direction of the original set. However, the rate formula in Eq.(7) is invariant under the  above operations as is apparent from the derivation of the method of images. We believe the reduction of the QED formulas to the simple form should be related to the above mentioned symmetries.

\section{Oscillations in the emission rate and closed-orbits}

We now analyze the spontaneous emission rate of an atom inside a $\pi/N$ wedge.
In Fig.3 we show the pattern of the spontaneous emission rate described by Eq.(7) for a $\theta=\pi/3$ wedge. The overall spontaneous emission rate pattern resembles ``an egg tray" turned upside
down.  There are two main features in the pattern in Fig.3. They are the suppression of the emission rate near the wedge surface and the regular oscillations of the rate inside the wedge. For the suppression of the rate, we note in the small $S$ limit, $G_{\|}(S)$ can be approximated by $(\frac{2}{3}-\frac{2}{15}S^2)$. Therefore the rate goes to zero quadratically as the emitting atom approaches the wedge surface. The quadratical nature of the rate suppression can be
seen in Fig.4 in which the emission rate is shown along two straight lines parallel to x-axis and y-axis respectively.

As we will explain next, the oscillations are related to some special photon closed-orbits along which the emitted photon travels away from and returns to the emitting atom inside the wedge.
The oscillations in the cross sections for photo-ionization and photo-detachment in the presence of external fields are successfully related to the closed-orbits of photo-electron going away and returning to the nucleus \cite{DuCOT1,DuCOT2,DuCOT3,DuCOT4,kleppnerdelos}.
Recently the oscillations in the spontaneous
emission rate have also been related to photon closed-orbits going away from and returning the emitting atom in two systems \cite{DuSE1,DuSE2}.
In the following, we will first analyze the oscillations in the emission rate for a $\pi/3$ wedge and will demonstrate the correspondence between the oscillations and some photon closed-orbits. A similar analysis is then performed for a $\pi/4$ wedge which confirms our understanding.

we define a
scaling variable $\alpha$ by $\rho=\alpha\lambda_0$, where $\lambda_0=2\pi/k$ is the wavelength in vacuum.
We now examine the spontaneous emission rate in the wedge as a function
of $\alpha$ with an arbitrary but fixed $\phi$ of the emitting atom.
Define the modified Fourier transform (FT)
\begin{equation}\label{7}
\widetilde{\gamma}(S)=\int_{\alpha_1}^{\alpha_2}[\gamma(\alpha)-\gamma_0]\alpha
e^{i\alpha S}d\alpha,
\end{equation}
where $\gamma(\alpha)$ represents the dependence of the emission rate as
a function of the scaling variable $\alpha$. In our numerical calculations, we took $\alpha_1=1.0$, $\alpha_2=26.0$ and the integration  step size $\Delta\alpha=0.01$.
As an example, for $\phi=\pi/9$, the emission rate as
a function of the scaling variable $\alpha$ for the $\theta=\pi/3$ wedge is shown in Fig.5(a). One can see the rate in the wedge oscillates around the free space rate and the oscillation amplitude is reduced as $\alpha$ increases or the emitting atom is further away from the wedge axis. The absolute value of the calculated modified Fourier transform is presented in Fig.5(b). There are clearly four
large peaks in Fig.5(b). These four peaks imply the emission rate in Fig.5(a) contains four main frequencies.
We now provide an explanation of the physical meaning for the peaks
and give a quantitative description for the peak positions and peak
heights.

We note when $S$ is larger than about $\pi$, $G_{\|}(S)=\frac{\sin S}{S}$ is an accurate approximation of $G_{\|}(S)$. When this approximation is made in Eq. (7),
the emission rate inside the wedge takes the following form
\begin{equation}
\gamma(\alpha)=\gamma_{0}+\sum_{j}
(\frac{3\gamma_{0}}{2S^0_j\alpha})\sin(S^0_{j}\alpha+\varphi_{j}),
\end{equation}
which is in the form of closed-orbit theory for absorption spectra \cite{DuCOT1,DuCOT2,DuCOT3,DuCOT4}. Eq.(9) can be interpreted using  closed-orbit theory.
$\gamma_{0}$ is a background rate associated with the initial
emission process in which the emitted photon leaves the atom and
never returns. It is equal to the spontaneous emission rate in free
space. The sum in Eq.(9) is over all
closed-orbits of emitted photon going away from and returning back to
the emitting atom; the emitted photon travels in straight paths in the wedge, and it follows the laws of reflection
when hitting the metal boundary of the wedge.
$S^0_j=kl^0_j$ is the ``action" around the $j$th closed-orbit, $k$ is the wave number, and $l^0_j$ is the geometric length for the $j$th closed-orbit corresponding to the case $\alpha=1$ or the distance between the emitting atom and the wedge axis is just equal to one wavelength; since each reflection by the wedge surface produces a $\pi$ phase shift, the phase $\varphi_j=-m_j\pi$ where $m_j$
counts the number of reflections by the wedge surface; the
amplitude $(\frac{3\gamma_{0}}{2S^0_j\alpha})$ provides a measure of the intensity of the returning group of photon wave, and in the present case,  it is inversely proportional to the length of the orbit.
The peak positions in the Fourier transform of Eq.(8)
are therefore given by the action $S^0_j$ of photon closed-orbits with
corresponding peak heights proportional to $g_j/l_j^0$, where $g_j$ is a
degeneracy factor counting closed-orbits with identical contribution to the peak. Both $S_j^0$ and $l_j^0$ are evaluated for the
system size corresponding to $\alpha=1$ or $\rho=\lambda_0$.

For a general wedge, we must launch trajectories in all direction from the location of the emitting atom at $(\rho, \phi,z)$ to search for all closed-orbits which leave from and return to the emitting atom. Each trajectory follows a straight line until it reaches the wedge and is reflected.
When the trajectory returns back to the location of emitting atom, a
closed-orbit is then formed. The closed-orbits must be on the cross-section plane of the atom because of the propagation rules of trajectories. If a trajectory leaves the atom off the cross-section plane, it can not return to the atom to form a closed-orbit. For a wedge with an arbitrary opening angle, the number of closed-orbit can be large. But for a $\pi/N$ wedge,
we can find all the closed-orbits from the images constructed earlier.
We find there is a one to one correspondence between the images and the closed-orbits.

In Fig.6, we show this correspondence for a $\theta=\pi/3$ wedge. There are five images labeled from 1 to 5 clockwise. There are also five closed-orbits. The first segment of the closed-orbit is given by the straight line connecting the corresponding image and the source.  The geometrical length of the closed-orbit is equal to the distance between the atom and the image.  The first closed-orbit leaves the emitting atom downward, it returns to the atom after being reflected back by the lower wedge surface; the second closed-orbit is first reflected by the lower wedge surface and then by the upper wedge surface before it returns to the emitting atom; the third closed-orbit is first reflected by the lower wedge surface, it then travels perpendicular to the upper wedge surface, finally it retraces back to the emitting atom after being reflected by the upper wedge surface. Similar descriptions can be given to the fourth closed-orbit and the fifth closed-orbit. We also find closed-orbit two and four have the same shape but their propagation directions are opposite.

The scaled ``actions" of the five closed-orbits depend only on the emitting atom's cylindrical angle $\phi$ and they are
\begin{eqnarray}\label{9}
 \nonumber S_1^0&=&2k\lambda_0\sin(\phi)=4\pi\sin(\phi),\\\nonumber
 \nonumber S_2^0&=&2k\lambda_0\sin(\frac{\pi}{3})=4\pi\sin(\frac{\pi}{3}),\\
 \nonumber S_3^0&=&2k\lambda_0\sin (\frac{\pi}{3}+\phi)
 =4\pi\sin (\frac{\pi}{3}+\phi),\\
 \nonumber S_4^0&=&S_2^0,\\
  S_5^0&=&2k\lambda_0\sin(\frac{\pi}{3}-\phi)=4\pi\sin(\frac{\pi}{3}-\phi).
\end{eqnarray}

Since each reflection by the wedge induces a phase loss of $\pi$,
the phases corresponding to the five closed-orbits above are
\begin{eqnarray}\label{10}
 \nonumber \varphi_1&=&\varphi_5=-\pi,\\
 \nonumber \varphi_2&=&\varphi_{4}=-2\pi,\\
 \varphi_3&=&-3\pi.
\end{eqnarray}

\begin{table}
\caption{The four peak positions and heights numerically extracted using FT are compared with the analytic predications discussed in the text for a $\pi/3$ wedge. The analytic results are represented by the crosses in Fig.5(b). Peaks are labeled from left to right with increasing value of the scaled action. }
\begin{tabular}{c c c c c c}
\hline\hline
  ~ & ~ & \multicolumn{2}{c}{position} & \multicolumn{2}{c}{height} \\
  ~ &~~~ peak label~~~ &~~~ numerical~~~ & ~~~analytic ~~~ & ~~~numerical~~~ & ~~~analytic ~~~ \\
  ~ & $1^{st}$ & 4.30 & 4.30  & 4.37 & 4.37 \\
  ~ & $2^{nd}$ & 8.08 & 8.08 & 2.30 & 2.32 \\
  ~ & $3^{rd}$ & 10.89 & 10.88 & 3.42 & 3.45 \\
  ~ & $4^{th}$ & 12.40 & 12.38 & 1.50 & 1.52 \\
  \hline\hline
\end{tabular}
\end{table}

We have used the formula $hg_i/l_i^0$ to compute the $i^{th}$ peak height located at $S^0_j$ for the $\pi/3$ wedge. The results from the analytic formulas with $\phi=\pi/9$ are marked as crosses in Fig.5(b). The degeneracy factor $g_i$ is equal to 1 for the first, second and fourth peak because one closed-orbit corresponds to each of the three peaks;  for the third peak near $10.89$, because there are two closed-orbits making the same contributions in both the emission rate and the FT, the corresponding degeneracy factor is 2. For the purpose of comparing with the numerical results, the constant $h$ was determined by matching with the first peak extracted from FT. We can see that all the peak positions and heights are accurately described by the analytic formulas. In Fig.5(b), we also show the closed-orbits correlated with the peaks. A detailed comparison is made in Table I between the numerically extracted peak positions and heights and the analytic formulas. We find they are in good agreement.

To confirm our understanding, we also studied the emission rate in a $\pi/4$ wedge setting $\phi=\pi/12$. The results are summarized and presented in Fig.7 and Fig.8. The images are constructed in Fig.7(a). The pattern of the emission rate inside the wedge calculated using Eq.(7) with $N=4$ is shown in Fig.7(b). This pattern also resembles ``an egg tray" turned upside
down. The seven images and the seven corresponding closed-orbits are illustrated in Fig.7(c).
The emission rate as a function of the scaling variable $\alpha$ is shown in Fig.8(a). The numerically calculated FT is displayed in Fig.8(b). We find there are six main peaks in the FT. The crosses near the peaks mark the results of the analytic formulas for the six peak positions and peak heights. The closed-orbits corresponding to the peaks are illustrated by the insert figures. From left to right, each and every peak is associated with one closed-orbit except the third one. The third peak corresponds to two closed-orbits. For the $\pi/4$ wedge, the numerically extracted peak positions and peak heights are listed together with the results of the analytic formulas in Table II. Again we find the numerical results and the analytic formulas agree well.

\begin{table}
\caption{The six peak positions and heights numerically extracted using FT are compared with the analytic predications discussed in the text for a $\pi/4$ wedge setting $\phi=\pi/12$. The analytic results are also represented by the crosses in Fig.8(b). Peaks are labeled from left to right with increasing value of the scaled action.  }
\begin{tabular}{c c c c c c}
\hline\hline
  ~ & ~ & \multicolumn{2}{c}{position} & \multicolumn{2}{c}{height} \\
  ~ &~~~ peak label~~~ &~~~ numerical~~~ & ~~~analytic~~~ &~~~ numerical~~~ &~~~ analytic~~~ \\
  ~ & $1^{st}$  & 3.24  &    3.25  &   5.82  &  5.79 \\
  ~ & $2^{nd}$  & 6.30  &  6.28   &  3.05  &  3.00 \\
  ~ & $3^{rd}$  &8.88   &  8.89   &  4.37  &   4.24 \\
  ~ & $4^{th}$&10.85  &   10.88  &   1.67 & 1.73 \\
  ~ & $5^{th}$ & 12.13 &  12.14  &  1.53  &  1.55 \\
  ~ & $6^{th}$ &12.56  &   12.57   &  1.45  &  1.50 \\
  \hline\hline
\end{tabular}
\end{table}

\section{ Conclusions }
In conclusion, we have studied the spontaneous emission of an atom
in a $\pi/N$ wedge using a method of images. The method of images is particular simple and the results for the emission rate are consistent with the results of QED. From the method of images, we can see that the rotation and inversion symmetries should play important roles in the reduction of complicated QED rate formulas to simple forms when the wedge opening angle is $\pi/N$, where $N$ is an arbitrary positive integer. For a wedge with an opening angle, which is not equal to $\pi/N$, the closed-orbit theory \cite{DuSE1,DuSE2,DuCOT2,DuCOT3} suggests the emission rate can still be expressed in a form similar to equation (7) although the number of oscillatory terms can be large, and the corresponding closed-orbits will have to be searched numerically\cite{DuCOT3}.

We analyzed in detail the oscillations in the emission rate using scaling variables and Fourier transformation for the $N=3$ and $N=4$ cases as examples. We found only four oscillations in a $\pi/3$ wedge and only six oscillations in a $\pi/4$ wedge.
We found a direct correspondence between the oscillations in the emission rate and  closed-orbits. We interpreted the oscillations
as interferences between the outgoing and incoming photon
waves traveling along closed-orbits from the atom to the atom.
The closed-orbits can be constructed readily from the method of images. The method of images provides a complementary perspective on the spontaneous emission process for an atom in a wedge, keeping in mind the difference between the method of images and the QED method \cite{Meschede}. It is interesting to note that in an recent experiment, Fleet \emph{et al} has put a classical dipole radiating at wavelengths of the order of a few centimeters inside a wedge-shaped cavity and measured the radiated power\cite{Fleet}. They observed that the classical dipole also exhibit effects similar to the Purcell effects. However, the classical dipole antenna is not a point source and can not yet simulate the emission of an atom in a wedge. On the other hand, the recent observation of the signature of photon periodic orbits in the spontaneous emission spectra in laterally confined vertically emitted cavities \cite{Chen} provides a very encouraging sign that an observation of the photon closed-orbits in a wedge is possible in the near future.

\begin{center}
{\bf ACKNOWLEDGMENTS}
\end{center}
\vskip8pt This work was supported by NSFC grant Nos.
10804066 and 11074260, SXNSF grant No. 2009011004 and TYAL.


\newpage
\begin{figure}
\includegraphics[scale=0.8,angle=-0]{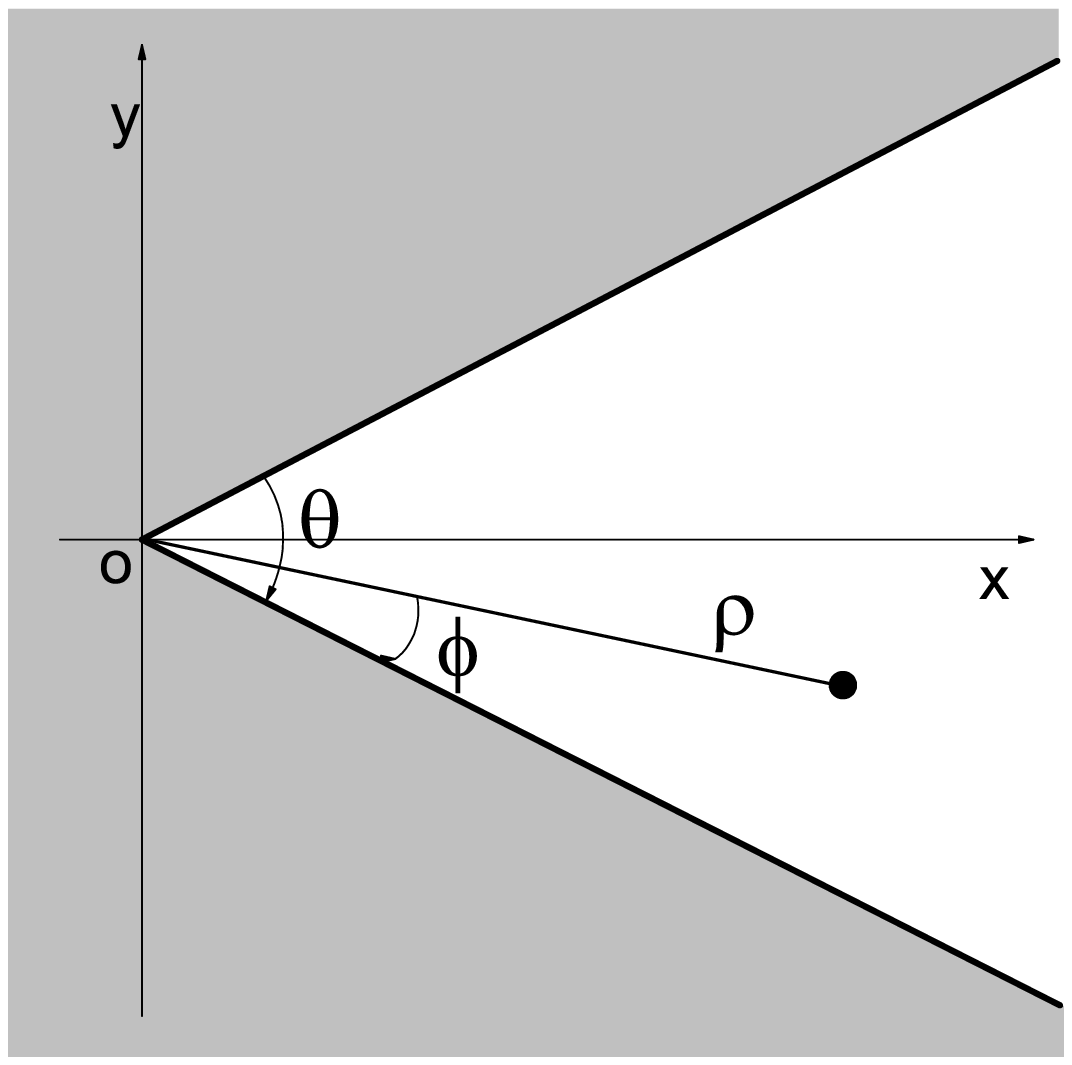}
\caption{Schematic diagram for an emitting atom (black dot) in a wedge with an opening angle $\theta=\pi/3$. Both cylindrical and Cartesian coordinate systems are used and illustrated. The z-axis is in the direction perpendicular to the paper. The x-axis is in the bisector plane of the wedge. }
\end{figure}

\begin{figure}
\includegraphics[scale=0.8,angle=-0]{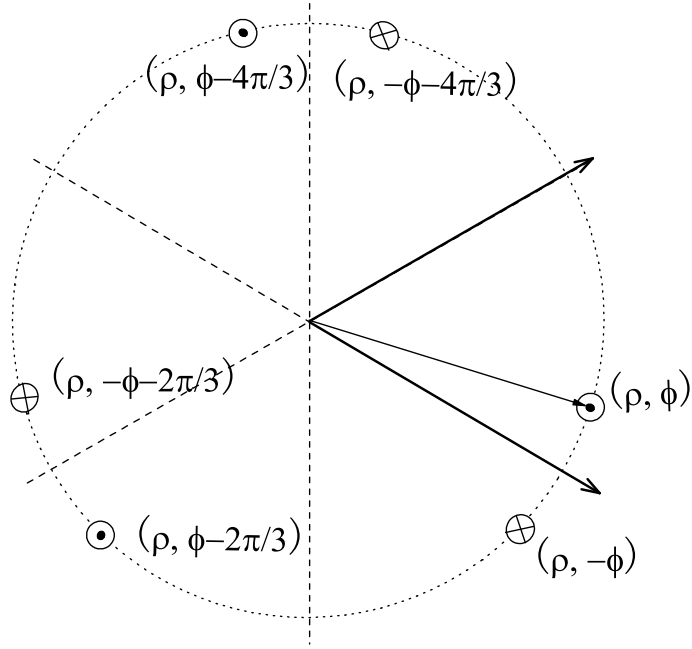}
\caption{The emitting dipole and their images for the $\theta=\pi/3$ wedge. The images represented by $\odot$ have dipoles identical to the  dipole of emitting atom while the images represented by $\otimes$ have opposite dipoles. }
\end{figure}

\begin{figure}
\includegraphics[scale=0.8,angle=-0]{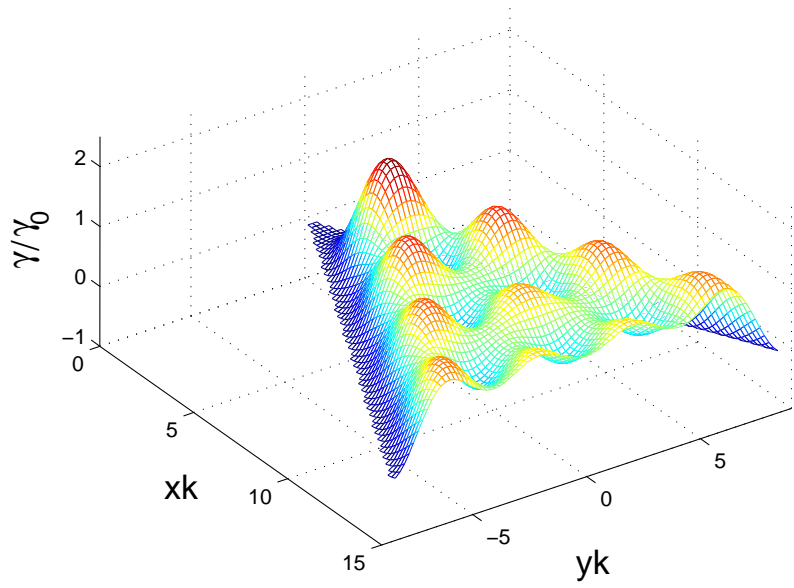}
\caption{The pattern of the spontaneous emission rate is formed as the position of the emitting atom is varied inside a wedge with an opening angle $\theta=\pi/3$.  $\gamma_0$ is the emission rate without the wedge. The pattern is like ``an egg tray" upside down.}
\end{figure}

\begin{figure}
\includegraphics[scale=0.8,angle=-0]{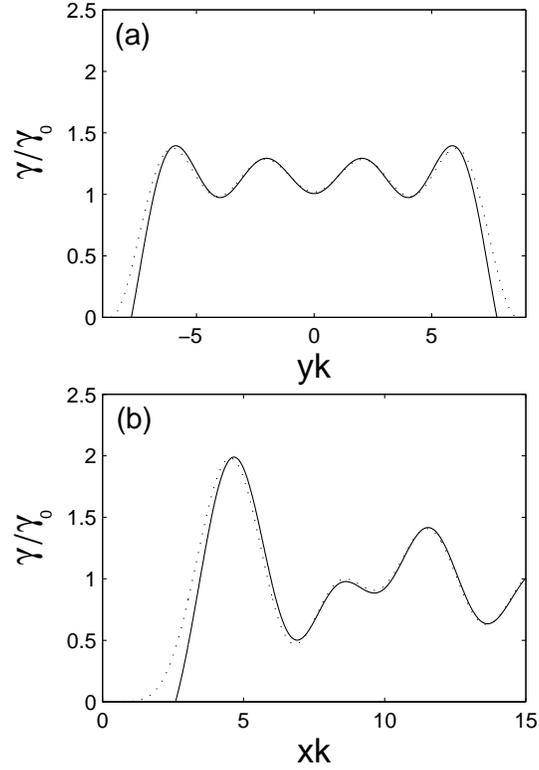}
\caption{The spontaneous emission rate for the emitting atom at (a) $kx=15$ and along the y-axis (b) $y=0$ and along the x-axis. Dotted lines correspond to Eq.(7) with $G_{\|}$ from Eq.(6) while solid lines correspond to Eq.(7) with $G_{\|}$ approximated by $\frac{\sin S}{S}$.  }
\end{figure}

\begin{figure}
\includegraphics[scale=0.7,angle=-0]{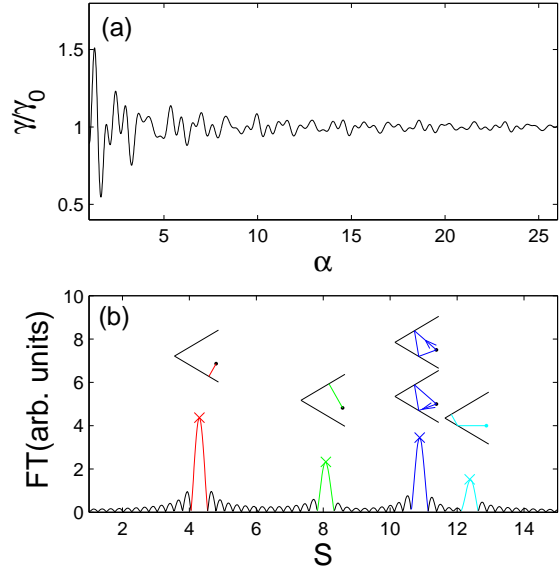}
\caption{(a) The spontaneous emission rate for the $\theta=\pi/3$ wedge as a function of the scaling parameter $\alpha=\rho/\lambda_0$ with $\phi=\pi/9$ for the emitting atom; (b) FT (absolute value) of the above emission rate as defined in Eq.(8). The crosses near the peaks mark the peak positions and heights obtained from the analytic formulas discussed in the text. The closed-orbits corresponding to the peaks are also displayed and illustrated with the same colors (online). }
\end{figure}

\begin{figure}
\includegraphics[scale=1.0,angle=-0]{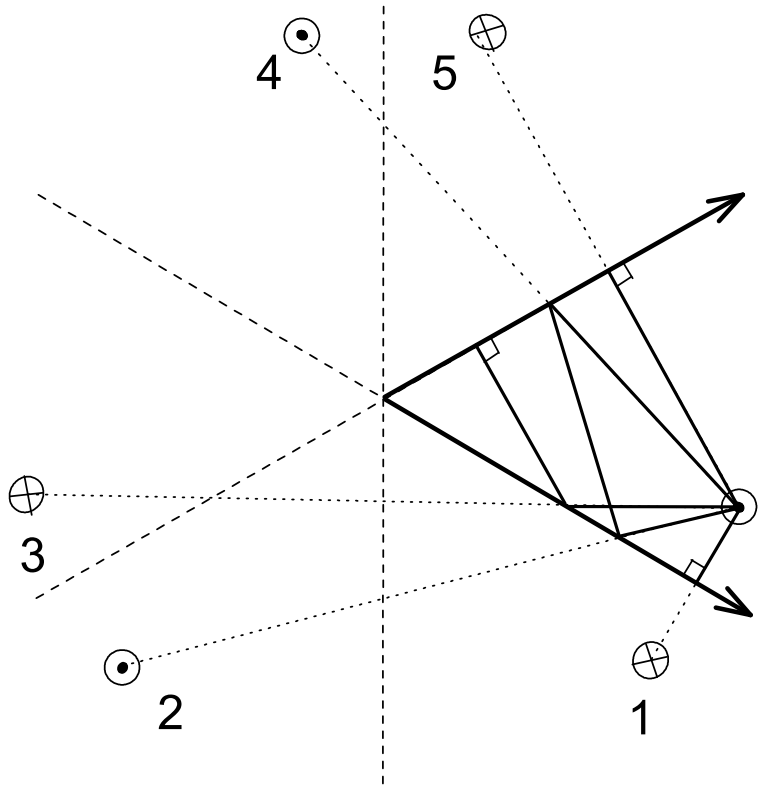}
\caption{The closed-orbits and their corresponding images in a $\pi/3$ wedge.
Each closed-orbit corresponds to one image. The first segment of the closed-orbit is given by the straight line connecting the corresponding image and the source. }
\end{figure}

\begin{figure}
\includegraphics[scale=0.7,angle=-0]{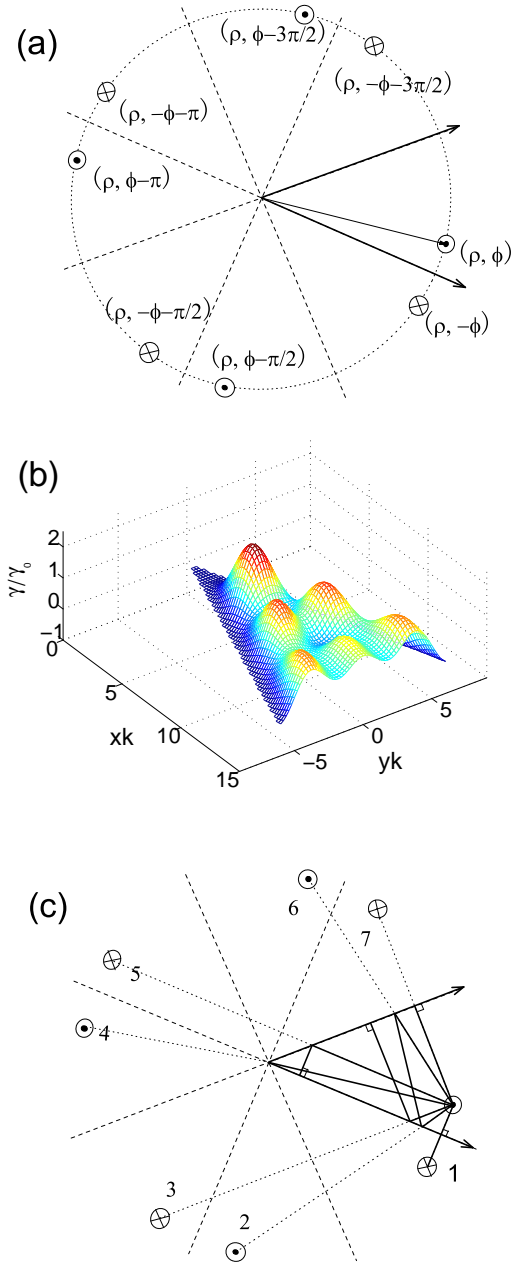}
\caption{The images,the emission rate and the closed-orbits for a $\pi/4$ wedge.
(a) The images are constructed according to the method discussed in the text;  (b) the pattern of the emission rate inside the wedge as a function of the position of the emitting atom; (c) the closed-orbits are constructed from their corresponding images. The first segment of each closed-orbit is given by the straight line connecting the corresponding image and the source. }
\end{figure}

\begin{figure}
\includegraphics[scale=0.8,angle=-0]{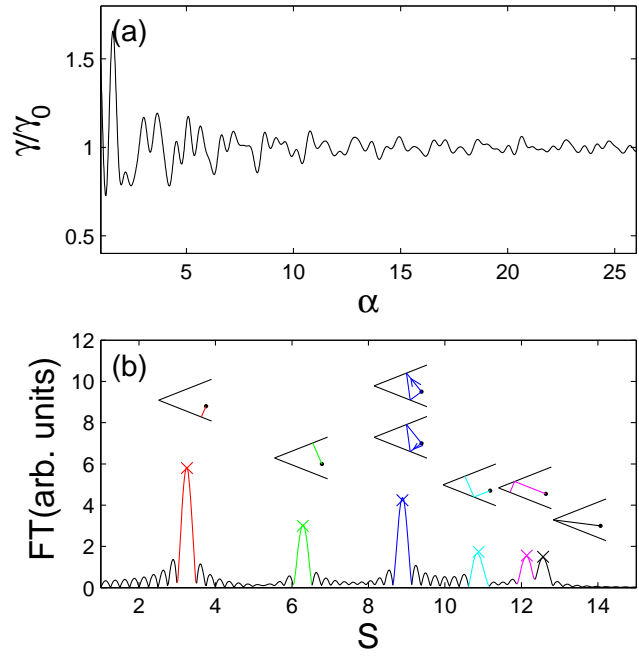}
\caption{Similar to Fig.5 but for the $\pi/4$ wedge with $\phi=\pi/12$.  }
\end{figure}


\end{document}